\documentclass[twocolumn,10pt,superscriptaddress,times]{revtex4-2}
\usepackage{graphics,graphicx,epsfig,ulem,epstopdf,bm,longtable,url,geometry,datetime,physics}
\usepackage[colorlinks=true,linkcolor=blue,urlcolor=blue,citecolor=blue]{hyperref}
\usepackage{amsmath,amssymb,latexsym,float,amsfonts,hyperref,color,setspace,overpic}
\usepackage{microtype,comment} 
\usepackage[caption=false]{subfig} 
\usepackage[ansinew]{inputenc}
\usepackage[usenames,dvipsnames]{pstricks}
\usepackage[pagewise,mathlines]{lineno}

\def\footnoterule{\kern -1mm \hrule width 6.0cm \kern 2.2mm}%
\usepackage{tikz,xcolor,hyperref}% Make Orcid icon
\definecolor{lime}{HTML}{A6CE39}
\DeclareRobustCommand{\orcidicon}{%
    \begin{tikzpicture}
    \draw[lime, fill=lime] (0,0)
    circle [radius=0.16]
    node[white] {{\fontfamily{qag}\selectfont \tiny ID}};\draw[white, fill=white] (-0.0625,0.095)
    circle [radius=0.007];
    \end{tikzpicture}
    \hspace{-2mm}}
\foreach \x in {A, ..., Z}
{\expandafter\xdef\csname orcid\x\endcsname{\noexpand\href{https://orcid.org/\csname orcidauthor\x\endcsname}{\noexpand\orcidicon}}}

\begin{document}
 %%%%%%%%%%%%%%%%%%%%%%%%%%%%%%%%%%%%%%%%%%%%%%%%%%%%%%%%%%%%
\title{Energy-Invariant Catalysis of Stable Ergotropy in Strongly Coupled Spin-Chain Quantum Batteries}

%{\small} {\footnotesize},{\scriptsize},{\tiny} µÈ£¬´óÐ¡ÒÀ´ÎµÝ¼õ
%\thanks{}
\author{Zi-Yi Peng$^{1}$}
\author{Shun-Cai Zhao\orcidA{}$^{1,2}$}
\email[Corresponding author: ]{zsczhao@126.com.}
\author{Liang Luo$^{2}$}
\author{Ni-Ya Zhuang$^{2}$}
\affiliation{$^{1}$Optoelectronic Information Science and Engineering, Faculty of Science, Kunming University of Science and Technology, Kunming, 650500, PR China}
\affiliation{$^{2}$Center for Quantum Materials and Computational Condensed Matter Physics, Faculty of Science, Kunming University of Science and Technology, Kunming, 650500, PR China}
%\affiliation{Department of Physics, Faculty of Science, Kunming University of Science and Technology, Kunming, 650500, PR China}

\begin{abstract}
Quantum batteries (QBs) provide a platform for exploring quantum-scale energy storage, yet most existing analyses rely on weak-coupling and Markovian approximations. In realistic implementations operating in strongly coupled non-Markovian regimes, environmental memory effects induce pronounced oscillations of the maximum extractable work (ergotropy), hindering stable energy output. Here, we investigate the stabilization of ergotropy in a spin-chain QB assisted by an energy-invariant catalyst, namely an auxiliary subsystem whose average energy remains unchanged during the evolution. 
The dynamics are described by a Nakajima¨CZwanzig master equation with a Gaussian memory kernel, enabling a systematic characterization of non-Markovian effects. 
Our results show that the memory-kernel parameters, the spin number, and the characteristic frequencies of both the cavity field and the local excitations jointly regulate the ergotropy dynamics.
Compared with the uncatalyzed case, the catalyst effectively reshapes the system energy spectrum, markedly suppresses non-Markovian oscillations, and promotes a quasi-stationary regime of extractable work. 
These findings provide a practical strategy for stabilizing energy flows in strongly coupled open quantum systems, offering theoretical guidance for the development of robust quantum energy devices and contributing to ongoing research in quantum thermodynamics.

\begin{description}
%\item[PACS]{ }
\item[Keywords]{Quantum batteries, Ergotropy stabilization, Physical catalysis, Non-Markovian \\ strong coupling }
\end{description}
\end{abstract}
\date{\today}%,~\currenttime}

\maketitle
%\tableofcontents
\section{Introduction}\label{sec:Introduction}
%\linenumbers

Understanding energy storage and transfer\cite{Campaioli2024} in open quantum systems is a central problem in quantum thermodynamics\cite{Ferraro2026}, with significant implications for nanoscale devices, quantum communication, and energy harvesting technologies\cite{Scully2010,Vinjanampathy2016,Goold2016,Quach2023}. Quantum batteries (QBs) have emerged as a promising class of quantum devices specifically designed to store and deliver energy by exploiting quantum mechanical principles ~\cite{Alicki2013,Binder2015,Ferraro2018,Campaioli2018,Bhattacharjee2021,Bhattacharyya2024}. Beyond their conceptual appeal, QBs are of practical relevance for realizing fast and efficient nanoscale energy storage~\cite{Andolina2019,Binder2015,Francica2017}, where quantum coherence and correlations can play an essential role~\cite{Gumberidze2019,Sen2021,Gumberidze2022}. A central question in this context is how QB performance-typically quantified by the ergotropy or charging power-is modified by interactions with external environments~\cite{Farina2019,Barra2019,Seah2021,Downing2024,Cavaliere2025}.

Most theoretical models of QBs assume weak system-environment coupling and employ the Markovian approximation \cite{Andolina2019}, allowing dissipation to be treated via memoryless Lindblad dynamics \cite{Le2018,Santos2019,Quach2020,Arrachea2023}. These models, while tractable, often neglect memory effects and fail to capture strong back-action from structured reservoirs, such as cavities or spin baths, which are prevalent in solid-state \cite{Ferraro2018,PhysRevA.90.022110} or circuit-QED implementations \cite{Quach2020,Hu2022}. Crucially, dissipative spin-cavity systems coupled to such structured reservoirs \cite{Lee2021,Hu2025} have been shown to sustain long-time nonclassicality and entanglement, features that are robustly sensitive to both coupling strength and memory-driven feedback \cite{MOHAMED20104115,Abdelghany2021}. These observations suggest that similar non-Markovian mechanisms could be strategically harnessed to suppress ergotropy decay and stabilize the performance of quantum batteries. Recent works have begun exploring non-Markovian dynamics \cite{Seah2021,Barra2019} or strong-coupling regimes \cite{Rossini2019,Andolina2019}, yet a systematic framework that treats both remains limited.

Another emerging concept in quantum thermodynamics is that of \textit{quantum catalysis}~\cite{PhysRevLett.85.437,Jonathan1999,Li2011,LipkaBartosik2021,Kondra2021a,Lie2023,RevModPhys.96.025005}. In this context, a catalyst is an auxiliary quantum system that assists a thermodynamic process without being consumed or fundamentally altered. Two main classes of quantum catalysis have been identified~\cite{Chaki2024universal,PhysRevA.107.042419}: \textit{state-invariant catalysis}, where the catalyst's quantum state remains unchanged during evolution, and \textit{energy-invariant catalysis}, where only the expectation value of the catalyst's energy remains constant\cite{Guryanova2016,Fellous2020}. While state-invariant catalysis is often employed in coherence-assisted work extraction protocols~\cite{PhysRevA.107.042419}, the energy-invariant variant is particularly relevant for open-system implementations, where full state preservation is experimentally challenging but energy constraints can be enforced.

Energy-invariant catalysis permits coherent or incoherent interactions between the catalyst and the system, provided the catalyst's average energy is conserved throughout the process. This definition aligns with scenarios in which the catalyst can be isolated or actively stabilized to maintain its energy, as in certain cavity-QED or solid-state architectures. Such catalysts can induce nontrivial system dynamics, enhancing ergotropy or work output while maintaining thermodynamic consistency. However, the operational role of energy-invariant catalysis under strong non-Markovian coupling-where environmental memory and back-action are significant-remains largely unexplored, particularly in relation to stabilizing ergotropy in QBs.

In our previous work~\cite{Zhao2025nspin}, we investigated the ergotropy dynamics of an $N$-spin-chain QB thermally charged via a cavity beyond the Markovian regime, showing that strong coupling to a structured environment produces pronounced ergotropy oscillations that can be partially mitigated via coherent control or interaction tuning. The potential stabilizing role of auxiliary degrees of freedom-especially physical catalysts-was not addressed.

Here, we address this gap by investigating how a physically defined energy-invariant catalyst influences ergotropy dynamics in a spin-chain QB~\cite{Le2018} strongly coupled to a bosonic cavity. The system dynamics are modeled using a Nakajima--Zwanzig-type master equation~\cite{Smirne2010,Xia2024} with a structured Gaussian memory kernel~\cite{Ferialdi2016}, enabling us to capture strong-coupling non-Markovian effects. We show that tuning the  kernel parameters ($\kappa_1$, $\kappa_2$), spin number $N$, single-mode cavity field frequency $\omega_c$ and local excitation energy $\omega_a$, can suppress large-amplitude ergotropy oscillations and establish a quasi-stationary regime. These findings highlight energy-invariant catalysis as a viable control mechanism for stabilizing work output in realistic quantum thermodynamic platforms~\cite{Joshi2022,Konar2022}.

This paper is organized as follows. In Section of Model, we present the full Hamiltonian, system--reservoir structure, and catalyst coupling scheme. The non-Markovian master equation approach and numerical implementation are also described. The main results are reported inSection of results, where we compare the dynamics with and without physical catalysis and analyze the associated physical mechanisms. The conclusion concludes with a summary and outlook.

\section{Theoretical Model and Computational Methods}\label{sec:model}

We consider a QB model comprising a linear spin chain of $N$ identical two-level systems, collectively serving as the work repository, embedded in a cavity field and regulated by an auxiliary physical catalyst. The total system interacts with a bosonic cavity environment under strong coupling and non-Markovian dynamics.
While most previous QB models assume weak dissipation and Markovian approximations~\cite{Le2018,Santos2019}, our framework explicitly incorporates environmental memory effects through a non-local integral kernel. This enables us to capture significant back-action from structured reservoirs, which are increasingly relevant in circuit-QED~\cite{Quach2020,Ferraro2018} and solid-state platforms~\cite{PhysRevA.90.022110,Hu2022}.

\begin{figure}\label{fig1}
\center
\includegraphics[width=0.4\columnwidth]{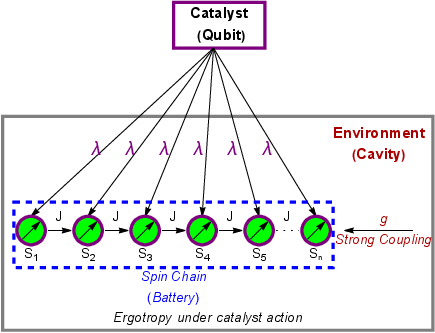 }\vspace{1cm}
\caption{(Color online) Schematics of a spin-chain quantum battery(QB) coupled to a cavity environment under the influence of a physical catalyst. The QB strongly interacting ($g$) with a structured cavity reservoir, consists of a chain of spin units {$S_i$ } coupled via exchange interaction $J$. A qubit-based catalyst, whose energy remains nearly constant during the evolution, is coupled to each spin via $\lambda$ . }
\end{figure}

The full system Hamiltonian is defined as (Here we set $\hbar$=$1$)
\begin{equation}
\hat{H}_{\text{tot}}=\hat{H}_{\text{photon}} + \hat{H}_{\text{spin}} + \hat{H}_J + \hat{H}_{\text{cat}} + \hat{H}_{\text{int}}^{(1)} + \hat{H}_{\text{int}}^{(2)},  \label{eq1}
\end{equation}
where $\hat{H}_\text{photon} $=$ \omega_c\, \hat{a}^\dagger \hat{a}$~represents the single-mode cavity field with frequency $\omega_c$, and the QB Hamiltonian is
\begin{equation}
\hat{H}_\text{spin}=\omega_a \sum_{i=1}^{N} \hat{\sigma}_i^+ \hat{\sigma}_i^-,%\qquad
\hat{H}_J=J \sum_{i=1}^{N-1} (\hat{\sigma}_i^+ \hat{\sigma}_{i+1}^- +\hat{ \sigma}_{i+1}^+ \hat{\sigma}_i^-), \label{eq2}
\end{equation}
with $\omega_a$ the local spin excitation energy and $J$ the nearest-neighbor exchange interaction strength. The cavity-spin coupling term is given by
\begin{equation}
\hat{H}_{\text{int}}^{(1)}=g \sum_{i=1}^{N} \left( \hat{a}^\dagger \hat{\sigma}_i^- + \hat{\sigma}_i^+ \hat{a} \right), \label{eq3}
\end{equation}
with coupling strength $g$ describing collective Jaynes-Cummings-type interactions. In thermodynamic resource theory, energy-invariant catalysis refers to transformations in which the catalyst's average energy remains constant, consistent with the framework in Refs.~\cite{Brandao2015,Lostaglio2015}. A catalyst, implemented here as a two-level qubit (or optionally a harmonic oscillator), couples to all spin sites with Hamiltonian
\begin{equation}
\hat{H}_\text{cat} = \frac{\omega_\text{cat}}{2} \hat{\sigma}_z, \qquad
\hat{H}_\text{int}^{(2)} = \lambda \sum_{i=1}^{N} (\hat{c}^\dagger \hat{\sigma}_i^- + \hat{\sigma}_i^+ \hat{c}),  \label{eq4}
\end{equation}
where $\hat{c}$ denotes the annihilation operator of the catalyst, and $\lambda$ is the catalyst-spin coupling strength.

We construct local operators via Kronecker embeddings. For instance, the cavity annihilation operator $\hat{a}_\text{ph}$ is embedded into the full space using
\begin{equation}
\hat{a} = \mathbb{\hat{I}}_\text{spin} \otimes \mathbb{\hat{I}}_\text{cat} \otimes \hat{a}_\text{ph},        \label{eq8}
\end{equation}
and similarly for the spin lowering operators $\hat{\sigma}_i^-$ and catalyst operator $\hat{c}$. This modular construction enables systematic evaluation of commutators and interaction terms.

The initial condition employed in our simulations corresponds to a fully charged battery, vacuum cavity, and ground-state catalyst, namely:
\begin{equation}
\rho(0)=\ket{0}_\text{photon} \bra{0} \otimes \ket{1\cdots1}_\text{spin} \bra{1\cdots1} \otimes \ket{0}_\text{cat} \bra{0}, \label{eq9}
\end{equation}

\noindent The initial state defined in Eq.\eqref{eq9} is not the vacuum of all subsystems, but a product state consisting of a \textit{vacuum cavity} ($|0\rangle_{\text{photon}}\langle0|$), a \textit{fully excited spin chain} ($|1\cdots1\rangle_{\text{spin}}\langle1\cdots1|$), and a \textit{ground-state catalyst} ($|0\rangle_{\text{cat}}\langle0|$). This choice models the experimentally relevant situation where the battery qubits are initialized by $\pi$ pulses\cite{PhysRevApplied.20.044077}. Additionally,  the single-subsystem Hamiltonian terms ($\hat{H}_{\text{photon}}$, $\hat{H}_{\text{spin}}$, $\hat{H}_{\text{cat}}$) individually conserve the excitation number of their respective subsystems, the cross-subsystem coupling terms-i.e., the cavity-spin interaction $\hat{H}_{\text{int}}^{(1)}$ (Eq.\eqref{eq3}) and catalyst-spin interaction $\hat{H}_{\text{int}}^{(2)}$ (Eq. \eqref{eq4})-explicitly break global excitation number conservation. This broken conservation enables energy transfer between subsystems (e.g., from spins to cavity photons or catalyst), laying the foundation for non-trivial ergotropy dynamics.

The state of the total system evolves under a non-Markovian master equation with an environment-induced memory kernel. Specifically, we adopt the Nakajima-Zwanzig (NZ) equation\cite{Smirne2010,Xia2024}
\begin{equation}
\frac{d\hat{\rho}(t)}{dt} = -i [\hat{H}_\text{tot}, \hat{\rho}(t)] + \int_0^t \Gamma(t,s)\, \mathcal{D}[\hat{\rho}(s)]\, ds.\label{eq5}
\end{equation}
%where $\mathcal{D}[\hat{\rho}]$= $(a \,\hat{\rho}\, \hat{a}^{\dagger} - \tfrac{1}{2} \{\hat{a}^{\dagger} \hat{a} \,\hat{\rho} ,~ \hat{\rho} \, \hat{a}^{\dagger} \hat{a} \}$
%denotes the dissipator associated with photon losses, modeled in Lindblad-like form but integrated over time. In this work, we focus on the Gaussian-type kernel\cite{Ferialdi2016}

For clarity, we expand the dissipator $\mathcal{D}[\hat{\rho}(s)]$ in Eq.\eqref{eq5} (accounting for photon loss from the cavity) into its explicit form:
\begin{equation}
\mathcal{D}[\hat{\rho}(s)] = \hat{a}_\text{ph}\hat{\rho}(s)\hat{a}^{\dagger}_\text{ph} - \frac{1}{2}\left\{ \hat{a}^{\dagger}_\text{ph}\hat{a}_\text{ph}, \hat{\rho}(s) \right\}, \label{eqaa}
\end{equation}
where $\hat{a}_\text{ph}$ and $\hat{a}^{\dagger}_\text{ph}$ are the cavity photon annihilation and creation operators (defined in Eq.\eqref{eq8}), and $\{ \cdot, \cdot \}$ denotes the anti-commutator. To simulate the time evolution of the battery-catalyst-cavity system under strong non-Markovian coupling, we numerically solve the Nakajima--Zwanzig master equation given in Eq.~(\ref{eq5}). The evolution is implemented in a truncated Hilbert space of total dimension $\dim(\mathcal{H}) = d_\text{photon} \times 2^N \times d_\text{cat}$, where $d_\text{photon}$ and $d_\text{cat}$ denote the photon and catalyst dimensions, respectively.

The corresponding \textit{collapse operator} for photon loss is derived by incorporating the Gaussian memory kernel $\Gamma(t,s)$(Please refer to Appendix A \ref{Appendix A}.):
\begin{equation}
\hat{C} = \sqrt{\Gamma(t,s)} \hat{a},                                  \label{eqbb}
\end{equation}
where
\begin{equation}
\Gamma(t,s) = \kappa_1 \exp[-\kappa_2 (t - s)^2],  \label{eq6}
\end{equation}
\noindent modulates the photon loss rate. This kernel captures non-Markovian environmental memory effects with tunable strength and width via $\kappa_1$ and $\kappa_2$, as observed in structured photonic reservoirs~\cite{Ferialdi2016}.
Although a Gaussian memory kernel is adopted here as a representative model of a structured reservoir with finite correlation time, the stabilization mechanism revealed below does not depend on its specific functional form. The essential ingredient is the existence of finite-time bath correlations, which enable transient energy backflow and partially counteract irreversible dissipation. 
More general non-Markovian environments-such as those associated with Lorentzian spectra, band-gap structures, or colored non-Gaussian noise-lead to different kernel profiles (e.g., exponential or oscillatory decay), but preserve the same qualitative mechanism as long as the bath correlation function deviates from the Markovian $\delta(t-s)$ limit. Therefore, the thermodynamic interpretation developed in this work extends beyond the Gaussian model to a broader class of experimentally relevant structured reservoirs.

%\textit{Generality beyond the Gaussian memory model.}
%Although the present analysis adopts a Gaussian memory kernel to model a structured reservoir with finite correlation time, the stabilization mechanism identified here does not rely on the specific Gaussian functional form. The essential physical ingredient is the presence of non-vanishing bath correlations over a finite temporal window, which enable transient energy backflow and partial protection of non-passive states.
%More generally, non-Markovian environments characterized by Lorentzian spectra, photonic band-gap structures, or colored non-Gaussian noise lead to memory kernels with exponential, oscillatory, or multi-timescale decay profiles. While these different kernels may quantitatively modify the memory amplitude and correlation time, the qualitative mechanism remains unchanged: whenever the bath correlation function deviates from the Markovian $\delta(t-s)$ limit, structured environmental feedback can temporarily counteract irreversible dissipation and reshape the ergotropy dynamics.
%In this sense, the Gaussian model should be regarded as a representative realization of finite-bandwidth reservoirs rather than a restrictive assumption. The thermodynamic interpretation developed in this work¡ªlinking energy backflow to enhanced stabilization¡ªtherefore extends to a broader class of experimentally relevant non-Markovian environments.

Importantly, we introduce the concept of a \textit{physical catalyst} with an invariant energy expectation value, in contrast to catalytic schemes that preserve the catalyst state itself~\cite{Rodriguez2023}. Our setting realizes catalytic behavior by maintaining the catalyst's energy stable throughout the evolution while enabling work extraction from the battery. This approach aligns with recent efforts to define operationally meaningful catalytic behavior in open quantum thermodynamics\cite{LipkaBartosik2021}.  To monitor the catalyst behavior, we also compute its instantaneous energy as
\begin{equation}
E_{\mathrm{cat}}(t) = \operatorname{Tr}[\hat{\rho}(t) \hat{H}_{\mathrm{cat}}],                   \label{eq10}
\end{equation}
which allows us to verify that the catalyst retains approximate energy-invariance throughout the evolution.

To assess the battery's performance, we focus on the \textit{ergotropy}~\cite{Allahverdyan2004,Frey2014,Brown2016,Campaioli2018}, which quantifies the maximum extractable work from a quantum state via unitary operations. For a given instantaneous density matrix $\rho(t)$, the ergotropy is defined as~\cite{Allahverdyan2004}
\begin{equation}
\mathcal{W}(t) = \text{Tr}[\hat{\rho}(t) \hat{H}_\text{B}] - \text{Tr}[\hat{\rho}_\text{passive}(t) \hat{H}_\text{B}],  \label{eq7}
\end{equation}
where $\hat{H}_\text{B} = \hat{H}_\text{spin} + \hat{H}_J$ is the battery Hamiltonian and $\hat{\rho}_\text{passive}$ is the passive state obtained by rearranging $\hat{\rho}$'s eigenvalues in decreasing order with respect to the energy eigenbasis~\cite{Pusz1978,Allahverdyan2004,Frey2014,Brown2016,Campaioli2018,Skrzypczyk2015,Llobet2015,Ghosh2020}. At each time $t$, we evaluate the battery energy $E_\text{B}(t) = \Tr[\rho(t)\, H_\text{B}]$ and the corresponding ergotropy $\mathcal{W}(t)$ by diagonalizing $\rho(t)$ and computing its passive counterpart~\cite{Allahverdyan2004}.
This quantity serves as our principal figure of merit.

Next, we numerically solve the NZ master equation using a truncated Hilbert space constructed via local operator embedding in the combined Hilbert space $\mathcal{H}_\text{photon} \otimes \mathcal{H}_\text{spin}^{\otimes N} \otimes \mathcal{H}_\text{cat}$, and convergence is verified by varying time-step tolerance and Hilbert space truncation. All simulations begin with a product state and track both the QB energy and its ergotropy under varying system parameters.

\section{Results and Discussion}\label{sec:results}

By numerically solving the Nakajima-Zwanzig master equation\eqref{eq5} under the specified memory kernel, we obtain the time-evolution of the elements of the density matrix $\hat{\rho}(t)$ for the quantum battery system. These density matrix elements encapsulate the complete information of the system's dynamical state, and are subsequently employed to determine the time-dependent ergotropy $\mathcal{W}(t)$, which establishes the link between the non-Markovian dynamical equations and the thermodynamic performance of the battery.

\begin{figure}[htbp]
\center
\includegraphics[width=0.25\columnwidth]{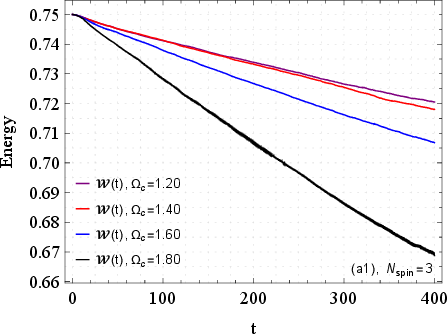 }\includegraphics[width=0.25\columnwidth]{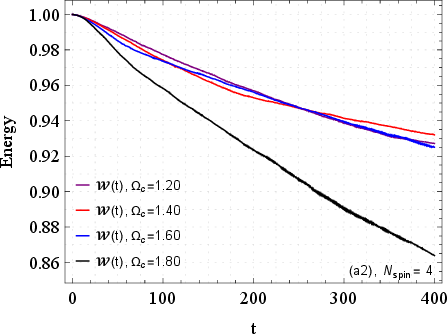 }\includegraphics[width=0.25\columnwidth]{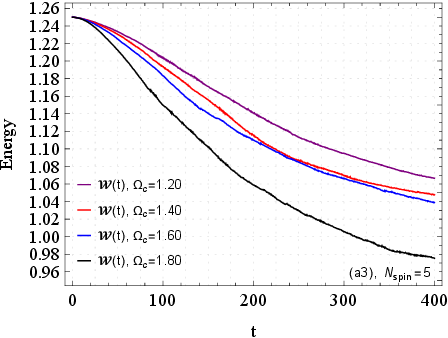 }\includegraphics[width=0.25\columnwidth]{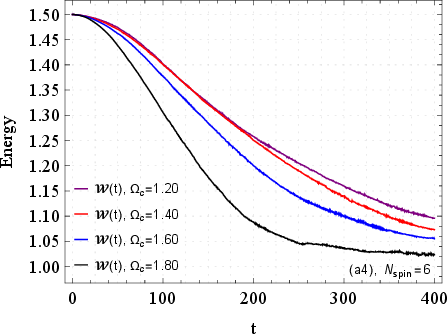 }
\includegraphics[width=0.25\columnwidth]{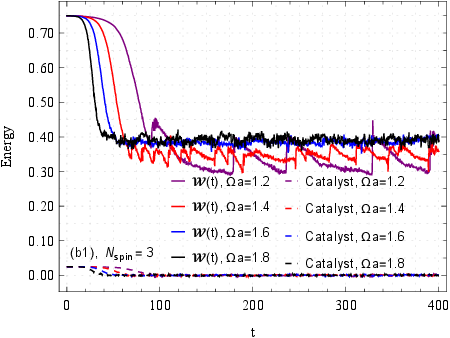 }\includegraphics[width=0.25\columnwidth]{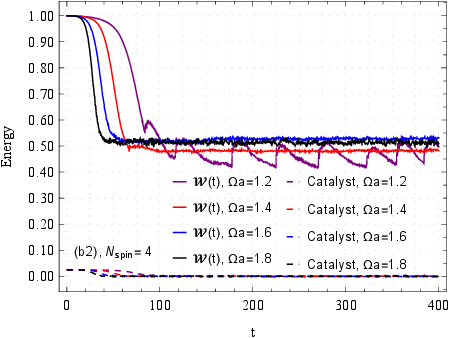 }\includegraphics[width=0.25\columnwidth]{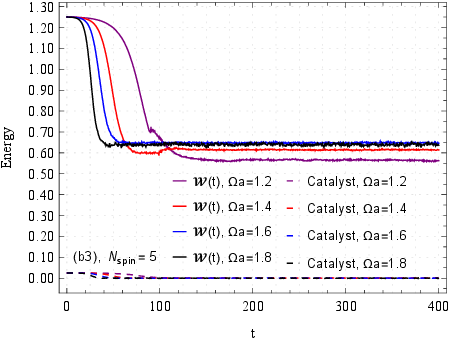 }\includegraphics[width=0.25\columnwidth]{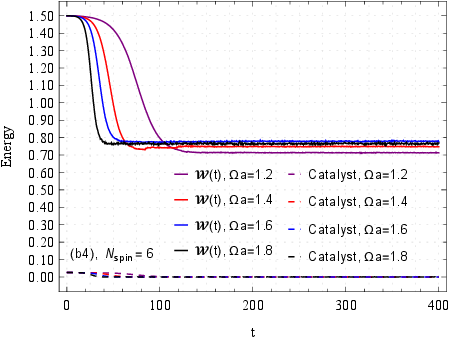 }
\caption{{\scriptsize(Color online) a(i)$_{(i=1\sim4)}$ Ergotropy dynamics for the catalyst-free battery-cavity system. b(i)$_{(i=1\sim4)}$ Ergotropy (solid line) and catalyst energy (dashed line) dynamics for the full battery-catalyst-cavity system at increasing single-mode cavity field frequency $\omega_c$. Simulation parameters are $\omega_a$=0.25, $\kappa_1$=1.2, $\kappa_2$=0.8, $\omega_{cat}$=0.05, $g$=$J$=$\lambda$=0.01.}}
\label{fig2}
\end{figure}

We now present the simulation results for the time evolution of ergotropy under different configurations of the battery-catalyst-cavity system. Our analysis is structured into two regimes. In the first, we demonstrate that physical catalysis stabilizes ergotropy dynamics within favorable control regimes, even when the negative regulatory parameter is suppressed. In the second, we reveal that under positively regulated conditions, the incorporation of a physical catalyst accelerates the stabilization of ergotropy output in the quantum battery, signifying the catalyst's transition from a passive energetic stabilizer to an active participant in the energy exchange process. For clarity, a unified illustration convention is adopted for all subsequent figures (Fig.(\ref{fig2})$\sim$ Fig.(\ref{fig5})): the first row depicts ergotropy dynamics without physical catalysts, while the second row (including dashed-line plots) depicts ergotropy dynamics with physical catalysts incorporated.

\subsection{Catalytic Stabilization under Negative Regulation}

To establish the catalytic role, we first fix the catalyst as a qubit with energy $\omega_\mathrm{cat}$=$ 0.05$, significantly smaller than the spin energy $\omega_a $=$ 0.25$. This configuration ensures it adheres to the physical definition of a catalyst-i.e., a low-energy auxiliary degree of freedom that induces nontrivial dynamics without dominating the energy landscape. The spin numbers $N$ in Fig.~\ref{fig2} are 3, 4, 5, and 6, respectively, labeled in each subfigure. Fig.~\ref{fig2}(a1)--(a4) exemplifies the ergotropy in the catalyst-free scenario. In contrast, Fig.~\ref{fig2}(b1)--(b4) exhibits ergotropy dynamics upon the introduction of the catalyst, as the single-mode cavity field frequency $\omega_c$ increases. In these panels, the nearly overlapping dashed lines represent the physical catalyst's energy, which remains effectively unchanged over the observed time window. This behavior illustrates that the catalyst acts as a low-energy auxiliary system, stabilizing the battery dynamics without significantly perturbing its own energy-consistent with the conceptual essence of a physical catalyst.

Fig.~\ref{fig2}(a1)$\sim$(a4) illustrates the time evolution of ergotropy under varying cavity-field frequencies $\omega_c$ and spin numbers $N$. The progressive attenuation of ergotropy with increasing $\omega_c$ reflects enhanced dissipative back-action (energy exchange with irreversible loss) between the spin ensemble and the cavity mode. As $\omega_c$ increases, this backreaction intensifies, leading to aggravated damping of ergotropy within the observed time window. Conversely, as the spin number $N$ increases, the system evolves toward a dynamically stabilized quasi-stationary ergotropy regime, with the most pronounced behavior at 
$\omega_c$=1.8 (Fig.~\ref{fig2}(a4)).

However, upon introducing physical catalysts, the progressive attenuation of ergotropy is effectively mitigated, as shown in Fig.~\ref{fig2}(b1)--(b4). Moreover, as the cavity-field frequency increases, quasi-stationary ergotropy is achieved at an earlier stage-evidenced by the black curves in each subfigure. We also note that increasing $N$ effectively suppresses ergotropy oscillations, as illustrated for $N$=3, 4 in  Fig.~\ref{fig2}(b2)) to (b3).

These results demonstrate that a physical catalyst fundamentally reshapes the energy-exchange landscape of the spin-cavity system. In the catalyst-free scenario [Fig.~\ref{fig2}(a1)--(a4)], the cavity field primarily induces dissipative back-action on the spin ensemble, leading to pronounced ergotropy attenuation and delayed stabilization. Here, energy flow is dominated by non-Markovian recurrences between spin and cavity modes, hindering steady energy extraction. In contrast, when the low-energy catalyst is incorporated [Fig.~\ref{fig2}(b1)--(b4)], dynamical interference modifies the dressed energy spectrum of the composite system, enabling facile hybridization of spin excitations with those in the battery, catalyst, and cavity subsystems. The catalyst reconfigures the available dynamical channels, fostering constructive interference between coherent and dissipative processes. This relieves backaction-induced coherence suppression and effectively suppresses non-Markovian recurrences. As a result, ergotropy attenuation is reduced, and the system reaches a quasi-stationary regime more rapidly. The catalyst thus acts as a passive yet functional controller, stabilizing energy flow without significant self-energy perturbation-validating its physical catalytic nature.

\begin{figure}[htbp]
\center
\includegraphics[width=0.25\columnwidth]{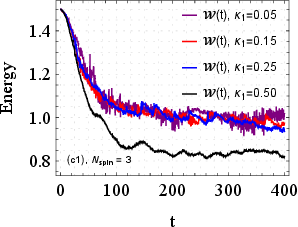 }\includegraphics[width=0.25\columnwidth]{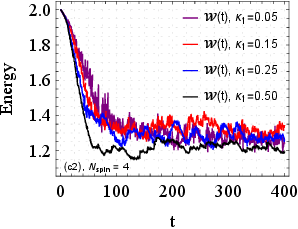 }\includegraphics[width=0.25\columnwidth]{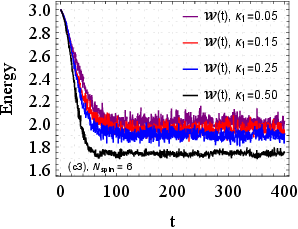 }\includegraphics[width=0.25\columnwidth]{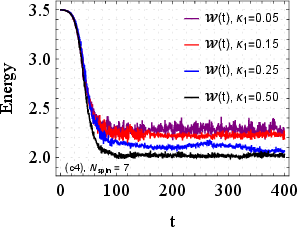 }
\includegraphics[width=0.25\columnwidth]{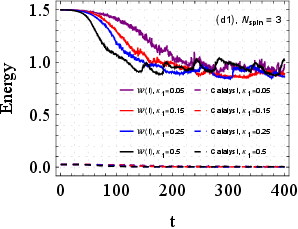 }\includegraphics[width=0.25\columnwidth]{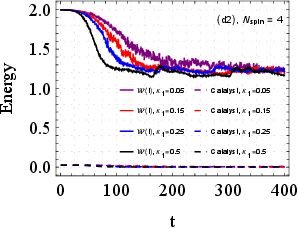 }\includegraphics[width=0.25\columnwidth]{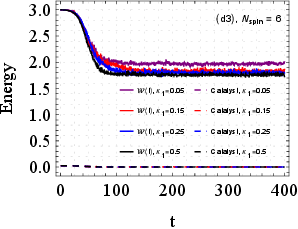 }\includegraphics[width=0.25\columnwidth]{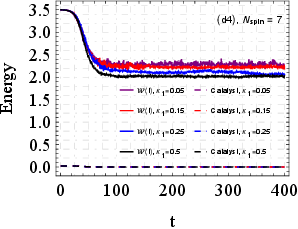 }
\caption{{\scriptsize(Color online) c(1)-c(4) Ergotropy dynamics for the catalyst-free battery-cavity system. d(1)-d(4) Ergotropy (solid lines) and catalyst energy (dashed lines) dynamics for the full battery-catalyst-cavity system, at increasing values of the memory kernel prefactor $\kappa_{1}$. Simulation parameters are $\omega_a$=0.5, $\omega_c$=0.25, $\lambda$=0.25, $\kappa_2$=0.05, $\omega_{cat}$=0.05, $g$=$J$=$\lambda$=0.05.}}
\label{fig3}
\end{figure}

To further dissect non-Markovian environmental memory effects on this behavior, Fig.~\ref{fig3} c(1)--c(4) analyzes $\mathcal{W}(t)$'s sensitivity to memory kernel prefactor $\kappa_{1}$ and spin number 
$N$ in the catalyst-free regime. As $\kappa_{1}$ increases, non-Markovian memory effects intensify, driving more energy to the environment. This reduces ergotropy oscillation amplitude, degrades its output, and gradually enhances system stability. For instance, the progressively narrowing distribution width of the purple, red, blue, and black curves in Fig.~\ref{fig3} c(3) along the horizontal axis indicates attenuated oscillations, while the black curve approaching the horizontal line at 1.75 corresponds to the minimum ergotropy output. Meanwhile, we also note that increasing the spin number $N$ effectively suppresses ergotropy oscillations, as illustrated for $N$=3, 4, 6, 7 in Fig.~\ref{fig3} c(1)--c(4). However, when the catalyst is present [Fig.~\ref{fig3} c(1)--c(4)], increasing $\kappa_{1}$ and $N$ instead lead to a reduction in oscillation amplitude, eventually driving the system into a near-steady ergotropy output regime.
 
\subsection{Catalytic Activation under Positive Regulation}

To further elaborate on the role of non-Markovian environmental memory effects, we now turn to the memory kernel width $\kappa_{2}$ in Eq.\eqref{eq6}, which, alongside $\kappa_{1}$, tunes the strength and temporal width of these memory effects. Specifically, $\kappa_{2}$ is anticipated to play a distinct role in shaping ergotropy output, and its influence will be dissected as follows.

\begin{figure}[htbp]
\center
\includegraphics[width=0.25\columnwidth]{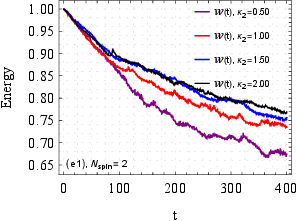 }\includegraphics[width=0.25\columnwidth]{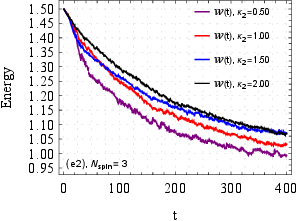 }\includegraphics[width=0.25\columnwidth]{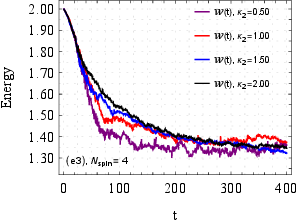 }\includegraphics[width=0.25\columnwidth]{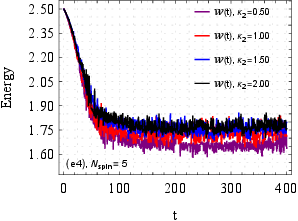 }
\includegraphics[width=0.25\columnwidth]{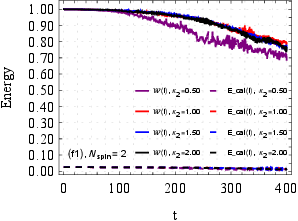 }\includegraphics[width=0.25\columnwidth]{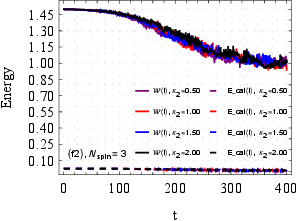 }\includegraphics[width=0.25\columnwidth]{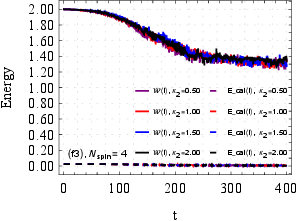 }\includegraphics[width=0.25\columnwidth]{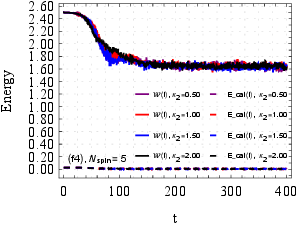 }
\caption{{\scriptsize(Color online)  e(i)$_{(i=1\sim4)}$ Ergotropy dynamics in the absence of catalysis, i.e., for the battery-cavity system alone, f(i)$_{(i=1\sim4)}$ Ergotropy (solid lines) and catalyst energy (dashed lines) dynamics for the full battery-catalyst-cavity setup at increasing the memory kernel width $\kappa_2$. Simulation parameters are $\kappa_1$=0.05, $\omega_a$=0.5, $\omega_c$=0.25, $\omega_{cat}$=0.05, $g$=$J$=0.01, $\lambda$=0.2.}}
\label{fig4}
\end{figure}

Fig.~\ref{fig4} plots the ergotropy dynamics as a function of increasing $\kappa_2$ for different spin numbers $N$, with other parameters held fixed. As $\kappa_2$(=0.5, 1.0, 1.5,2.0) gradually increases, the decrease in ergotropy output begins to slow down. Conversely, as the spin number $N$ increases, oscillations intensify in the catalyst-free regime, as shown in Fig.~\ref{fig4} e(1)--e(4).
In contrast, in the presence of a physical catalyst, variations in the memory kernel width $\kappa_2$ exhibit almost no significant influence on ergotropy output-evidenced by the nearly overlapping curves in Fig.~\ref{fig4} f(1)--f(4).

The results in Fig.~\ref{fig4} illustrate that in the catalyst-free regime (Fig.~\ref{fig4} e(1)--e(4)), increasing the memory kernel width $\kappa_2$ broadens the non-Markovian environmental memory, weakening dissipative back-action and thus slowing the decay of ergotropy; meanwhile, larger spin numbers $N$ enhance spin ensemble coherence, intensifying ergotropy oscillations. In contrast, when a physical catalyst is present (Fig.~\ref{fig4} f(1)--f(4)), it hybridizes the energy spectra of the battery, spin, and cavity subsystems, effectively screening the system from  $\kappa_2$ variations and stabilizing ergotropy output, as evidenced by the nearly overlapping curves.

\begin{figure}[htbp]
\center
\includegraphics[width=0.25\columnwidth]{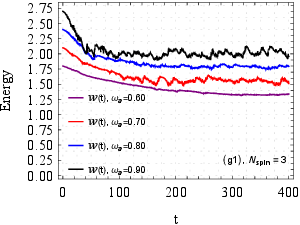 }\includegraphics[width=0.25\columnwidth]{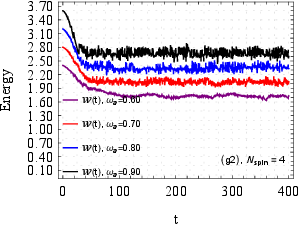 }\includegraphics[width=0.25\columnwidth]{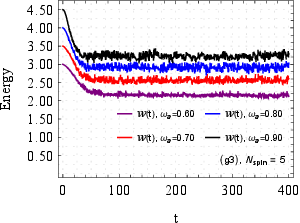 }\includegraphics[width=0.25\columnwidth]{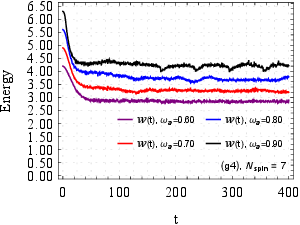 }
\includegraphics[width=0.25\columnwidth]{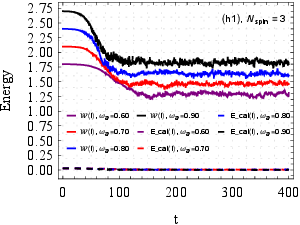 }\includegraphics[width=0.25\columnwidth]{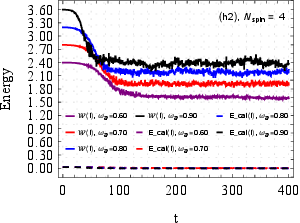 }\includegraphics[width=0.25\columnwidth]{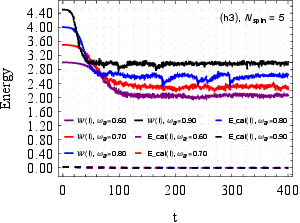 }\includegraphics[width=0.25\columnwidth]{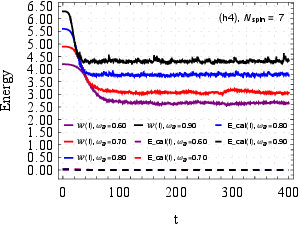 }
\caption{{\scriptsize(Color online) (a) Ergotropy dynamics in the absence of catalysis, i.e., for the battery-cavity system alone with  $g$=0.05, $J$=0.04. (b) Ergotropy (solid lines) and catalyst energy (dashed lines) dynamics for the full battery-catalyst-cavity setup at increasing the local spin excitation energy $\omega_a$ with  $g$=$J$=0.01. Simulation parameters are $\omega_c$=0.25, $\omega_{cat}$=0.05, $\kappa_1$=1.2, $\kappa_2$=1.8,  $\lambda$=0.01. }}
\label{fig5}
\end{figure}

Having elucidated the roles of memory kernel parameters and catalytic effects, we now turn to investigating the influence of the local spin excitation energy $\omega_a$ on ergotropy, as presented in Fig.~\ref{fig5}. This analysis aims to dissect how modulating $\omega_a$ reshapes the energy-level structure of the spin ensemble and its interplay with the battery-cavity system, both in the absence and presence of a physical catalyst. 

In the catalyst-free case [Fig.~\ref{fig5} g(1)--g(4)], increasing $\omega_a$ primarily alters the energy level structure of the spin ensemble, leading to a faster decay of ergotropy as the system's intrinsic frequency moves away from optimal resonance with the cavity mode. In the catalyst-free regime [Fig.~\ref{fig5} g(1)--g(4)], increasing $\omega_a$ induces a gradual rise in ergotropy output, albeit accompanied by increasingly pronounced oscillations [Fig.~\ref{fig5} g(1)--g(3)]. Conversely, for larger spin numbers ($N$=7), the oscillatory ergotropy output is effectively quenched [Fig.~\ref{fig5} g(4)]. 
Upon introducing a physical catalyst [Fig.~\ref{fig5} h(1)--h(4)], increasing $\omega_a$ solely enhances ergotropy output. Meanwhile, increasing spin numbers further boost the ergotropy output [Fig.~\ref{fig5} h(1)--h(4)]. The catalyst mediates coherent energy exchange between the battery and the spin-cavity subsystem; this catalytic interaction effectively compensates for the detuning induced by increasing 
$\omega_a$, thereby maintaining a more stable and robust ergotropy output--evidenced by the stabilized dynamics and suppressed oscillations, even for larger spin numbers.

\textit{Outlook} - This work demonstrates that energy-invariant catalysis, by hybridizing the system's energy spectrum and interfering with dissipative channels, can effectively stabilize ergotropy output against environmental memory effects and parameter variations. Future work could extend this concept in several directions. Theoretically, multi-catalyst configurations could be designed to engineer more robust spectral hybridization, while time-dependent control of catalyst coupling may further optimize dynamic interference. Incorporating finite-temperature effects would help assess catalytic performance under realistic noise. 
Experimentally, the predicted stabilization is accessible in platforms such as superconducting circuits and trapped-ion chains~\cite{Gu2017,Blatt2012,Blais2021}, where parameters like $g$, $\lambda$, $\kappa_{1}$, and $\kappa_{1}$ can be tuned to realize the regimes explored here. Beyond quantum batteries, the mechanism of spectrum-mediated stabilization~\cite{Joshi2022,Konar2022} may benefit quantum heat engines, light-harvesting complexes, and coherence protection in quantum networks.

\section{Conclusion}\label{sec:conclusion}

We have investigated the modulation of ergotropy in a non-Markovian spin-chain quantum battery mediated by an energy-invariant catalyst. Leveraging a Nakajima--Zwanzig-type non-Markovian master equation with a Gaussian memory kernel, we demonstrated that memory kernel parameters ($\kappa_1$, $\kappa_2$ ), spin number $N$,  single-mode cavity field frequency $\omega_c$ and local spin excitation energy $\omega_a$ synergistically govern ergotropy dynamics. The catalyst, by hybridizing the system's energy spectrum, suppresses deleterious oscillations and fosters a quasi-stationary ergotropy regime. This work uncovers a viable control strategy under pronounced environmental memory, advancing the understanding of non-Markovian open quantum systems and catalysis in quantum thermodynamics, and paving the way for robust quantum energy devices and coherent energy transfer systems.
 
Looking forward, this mechanism invites several promising extensions: (i) investigating multi-catalyst configurations or time-dependent catalysis schemes, (ii) generalizing to diverse spectral densities and higher-spin architectures, and (iii) experimental implementation in superconducting circuits or trapped-ion platforms with tunable structured environments. These avenues could enable quantum thermodynamic devices that harness environmental memory for enhanced performance rather than just mitigating its effects.

\section*{Data Availability Statement}

This manuscript has associated data in a data repository. [Authors' comment: All data included in this manuscript are available upon reasonable request by contacting with the corresponding author]. %The Supporting Information is available free of charge at: \href{https://github.com/zsczhao/Data-for-Non-Markovian-Spin-Chain-Quantum-Batteries} {https://github.com/zsczhao/Data-for-Non-Markovian-Spin-Chain-Quantum-Batteries}.

\section*{Acknowledgment}

This work is supported by the National Natural Science Foundation of China ( Grant Nos. 62065009 and 61565008 ).

\section*{ Appendix A}\label{Appendix A}

\noindent\textbf{1. Starting point: System-reservoir Hamiltonian}

We consider a generic system-bath Hamiltonian
\begin{equation}
\hat{H} = \hat{H}_S + \hat{H}_B + \hat{H}_{SB},
\end{equation}
with
\begin{equation}
\hat{H}_{SB} = \sum_k \left( g_k \hat{L} \, \hat{b}_k^\dagger + g_k^* \hat{L}^\dagger \hat{b}_k \right),
\end{equation}
where $\hat{L}$ is the system coupling operator (here proportional to $\hat{a}_{\mathrm{ph}}$), and $\hat{b}_k$ are bosonic bath operators.

\noindent\textbf{2. Nakajima-Zwanzig formalism}

Applying the standard projection operator technique and tracing over bath degrees of freedom up to second order in $\hat{H}_{SB}$ (Born approximation), one obtains the time-nonlocal master equation
\begin{equation}
\frac{d\hat{\rho}_S(t)}{dt}
=
-i[\hat{H}_S,\hat{\rho}_S(t)]
-
\int_0^t d s \, C(t-s)
\left[
\hat{L}, \hat{L}^\dagger(s-t)\hat{\rho}_S(s)
\right]
+ \mathrm{H.c.},
\end{equation}
where the bath correlation function is
\begin{equation}
C(t-s) = \langle \hat{B}(t) \hat{B}^\dagger(s) \rangle.
\end{equation}

\noindent\textbf{3. Correlation function from a structured reservoir}

For a bosonic environment initially in vacuum (or thermal equilibrium), the two-time correlation function reads
\begin{equation}
C(\tau) = \int_0^\infty d\omega \, J(\omega) e^{-i\omega \tau},
\end{equation}
where $J(\omega)$ is the spectral density of the reservoir.

We assume a structured photonic reservoir with a Gaussian spectral profile centered around $\omega_c$,
\begin{equation}
J(\omega) = \kappa_1
\exp\!\left[
-\frac{(\omega-\omega_c)^2}{2\sigma^2}
\right].
\end{equation}

\noindent\textbf{4. Emergence of the Gaussian memory kernel}

Substituting $J(\omega)$ into the Fourier transform, we obtain
\begin{equation}
C(\tau)
=
\kappa_1
\int_{-\infty}^{\infty}
d\omega \,
\exp\!\left[
-\frac{(\omega-\omega_c)^2}{2\sigma^2}
\right]
e^{-i\omega \tau}.
\end{equation}

Using the standard Gaussian integral identity,
\begin{equation}
\int_{-\infty}^{\infty}
e^{-a(x-b)^2} e^{-i x \tau} dx
=
\sqrt{\frac{\pi}{a}}
\exp\!\left[
-\frac{\tau^2}{4a}
- i b \tau
\right],
\end{equation}
we obtain
\begin{equation}
C(\tau)
=
\kappa_1'
\exp\!\left[
-\frac{\sigma^2 \tau^2}{2}
- i\omega_c \tau
\right].
\end{equation}

Neglecting the fast oscillating phase factor in the interaction picture (rotating-wave frame), the real envelope defines the memory kernel,
\begin{equation}
\Gamma(t,s)
=
\kappa_1
\exp\!\left[
-\kappa_2 (t-s)^2
\right],
\end{equation}
with $\kappa_2 = \sigma^2/2$ and $\kappa_1$ absorbing normalization constants.
The Gaussian kernel therefore emerges directly from assuming a reservoir with a Gaussian spectral density, corresponding physically to a structured cavity environment with finite bandwidth $\sigma$. The parameter $\kappa_1$ controls the overall coupling strength (memory amplitude), while $\kappa_2$ determines the memory time scale $\tau_c \sim 1/\sqrt{\kappa_2}$. In the limit $\kappa_2 \rightarrow \infty$, the kernel approaches $\delta(t-s)$ and the dynamics reduce to the Markovian Lindblad form.

\bibliography{ref_zsc}

@article{Campaioli2024,
  author  = {Campaioli, F. and Gherardini, S. and Quach, J. Q. and Polini, M. and Andolina, G. M.},
  title   = {Colloquium: Quantum batteries},
  journal = {Rev. Mod. Phys.},
  volume  = {96},
  number  = {},
  pages   = {031001},
  year    = {2024},
  doi     = {10.1103/RevModPhys.96.031001}
}

@article{Ferraro2026,
  author  = {Ferraro, D. and Cavaliere, F. and Genoni, M. G. and Benenti, G. and Sassetto, M.},
  title   = {Opportunities and challenges of quantum batteries},
  journal = {Nat. Rev. Phys.},
  volume  = {8},
  number  = {},
  pages   = {115--127},
  year    = {2026},
  doi     = {https://doi.org/10.1038/s42254-025-00906-5}
}

@article{Lee2021,
author = {Lee, C. H. and Lin, C. W.},
title = {A Two-Phase Fashion Apparel Detection Method Based on YOLOv4},
journal = {Appl. Sci.},
volume = {11},
number = {9},
pages = {3782},
year = {2021},
doi = {10.3390/app11093782}
}

@article{Hu2025,
author = {Hu, T. and Li, Y. and Hong, J. and Guo, D. and Li, X. and Chen, K.},
title = {Many-body localization properties of one-dimensional anisotropic spin-1/2 chains},
journal = {Quantum Inf. Process.},
volume = {24},
number = {7},
pages = {249},
year = {2025},
doi = {10.1007/s11128-025-04860-0}
}

@article{Quach2023,
author  = {Quach, J. Q. and Cerullo, G. and Virgili, T.},
title   = {Quantum batteries: the future of energy storage?},
journal = {Joule},
volume  = {7},
number  = {10},
pages   = {2195--2200},
year    = {2023},
doi     = {10.1016/j.joule.2023.09.003}
}

@article{MOHAMED20104115,
title = {Long-time death of nonclassicality of a cavity field interacting with a charge qubit and its own reservoir},
journal = {Phys. Lett. A},
author = { Mohamed, A. B. A.},
volume = {374},
number = {40},
pages = {4115-4119},
year = {2010},
doi = {https://doi.org/10.1016/j.physleta.2010.08.028}
}

@article{Abdelghany2021,
author = {Abdelghany, R. A. and Mohamed, A. B. A. and Tammam, M. and Kuo, Watson and Eleuch, H.},
title = {Tripartite entropic uncertainty relation under phase decoherence},
journal = {Sci. Rep.},
volume = {11},
number = {1},
pages = {11830},
year = {2021},
doi = {10.1038/s41598-021-90689-3}
}

@article{Goold2016,
  author = {Goold, J. and Huber, M. and Riera, A. and del Rio, L. and Skrzypczyk, P.},
  title = {The role of quantum information in thermodynamics--a topical review},
  journal = {J. Phys. A: Math. Theor.},
  volume = {49},
  number = {14},
  pages = {143001},
  year = {2016},
  doi = {10.1088/1751-8113/49/14/143001}
}

@article{PhysRevApplied.20.044077,
  title = {Active initialization experiment of a superconducting qubit using a quantum circuit refrigerator},
  author = {Yoshioka, T. and Mukai, H. and Tomonaga, A. and Takada, S. and Okazaki, Y. and Kaneko, N.-H. and Nakamura, S. and Tsai, J. S.},
  journal = {Phys. Rev. Appl.},
  volume = {20},
  issue = {4},
  pages = {044077},
  numpages = {11},
  year = {2023},
  publisher = {American Physical Society},
  doi = {10.1103/PhysRevApplied.20.044077},
  url = {https://link.aps.org/doi/10.1103/PhysRevApplied.20.044077}
}

@article{Brandao2015,
  author = {Brando, F. G. S. L. and Horodecki, M. and Ng, N. H. Y. and Oppenheim, J. and Wehner, S.},
  title = {The second laws of quantum thermodynamics},
  journal = {Proc. Natl. Acad. Sci. U.S.A.},
  volume = {112},
  number = {11},
  pages = {3275--3279},
  year = {2015},
  doi = {10.1073/pnas.1411728112}
}

@article{Lostaglio2015,
  author = {Lostaglio, M. and Jennings, D. and Rudolph, T.},
  title = {Description of quantum coherence in thermodynamic processes requires constraints beyond free energy},
  journal = {Nat. Commun.},
  volume = {6},
  pages = {6383},
  year = {2015},
  doi = {10.1038/ncomms7383}
}

@article{Vinjanampathy2016,
  author = {Vinjanampathy, S. and Anders, J.},
  title = {Quantum thermodynamics},
  journal = {Contemp. Phys.},
  volume = {57},
  number = {4},
  pages = {545--579},
  year = {2016},
  doi = {10.1080/00107514.2016.1201896}
}

@article{Scully2010,
  author = {Scully, M. O.},
  title = {Quantum Photocell: Using Quantum Coherence to Reduce Radiative Recombination and Increase Efficiency},
  journal = {Phys. Rev. Lett.},
  volume = {104},
  number = {20},
  pages = {207701},
  year = {2010},
  doi = {10.1103/PhysRevLett.104.207701}
}

@article{Guryanova2016,
  author = {Guryanova, Y. and Popescu, S. and Short, A. J. and Silva, R. and Skrzypczyk, P.},
  title = {Thermodynamics of quantum systems with multiple conserved quantities},
  journal = {Nat. Commun.},
  volume = {7},
  pages = {12049},
  year = {2016},
  doi = {10.1038/ncomms12049}
}

@article{Fellous2020,
  author = {Fellous-Asiani, S. and Alhambra, {\'A}. M. and Riera, A.},
  title = {Catalysis and coherence in thermodynamics},
  journal = {Quantum},
  volume = {4},
  pages = {272},
  year = {2020},
  doi = {10.22331/q-2020-06-29-272}
}

@article{Rodriguez2023,
  author = {Rodriguez, R. R. and Ahmadi, B. and Mazurek, P. and Barzanjeh, S. and Alicki, R. and Horodecki, P.},
  title = {Catalysis in charging quantum batteries},
  journal = {Phys. Rev. A},
  volume = {107},
  number = {4},
  pages = {042419},
  year = {2023},
  doi = {10.1103/PhysRevA.107.042419}
}

@article{Ferialdi2016,
  author = {Ferialdi, L.},
  title = {Exact closed master equation for Gaussian non-Markovian dynamics},
  journal = {Phys. Rev. Lett.},
  volume = {116},
  number = {12},
  pages = {120402},
  year = {2016},
  doi = {10.1103/PhysRevLett.116.120402}
}

@article{Smirne2010,
  author = {Smirne, A. and Vacchini, B.},
  title = {Nakajima-Zwanzig versus time-convolutionless master equation for the non-Markovian dynamics of a two-level system},
  journal = {Phys. Rev. A},
  volume = {82},
  number = {2},
  pages = {022110},
  year = {2010},
  doi = {10.1103/PhysRevA.82.022110}
}

@article{Xia2024,
  author = {Xia, Z. and Garcia-Nila, J. and Lidar, D. A.},
  title = {Markovian and Non-Markovian Master Equations versus an Exactly Solvable Model of a Qubit in a Cavity},
  journal = {Phys. Rev. Applied},
  volume = {22},
  number = {1},
  pages = {014028},
  year = {2024},
  doi = {10.1103/PhysRevApplied.22.014028}
}

@article{Joshi2022,
  author = {Joshi, J. and Mahesh, T. S.},
  title = {Experimental investigation of a quantum battery using star-topology NMR spin systems},
  journal = {Phys. Rev. A},
  volume = {106},
  number = {4},
  pages = {042601},
  year = {2022},
  doi = {10.1103/PhysRevA.106.042601}
}

@article{Konar2022,
  author = {Konar, T. K. and Lakkaraju, L. G. C. and Ghosh, S. and Sen(De), A.},
  title = {Quantum battery with ultracold atoms: Bosons versus fermions},
  journal = {Phys. Rev. A},
  volume = {106},
  number = {2},
  pages = {022618},
  year = {2022},
  doi = {10.1103/PhysRevA.106.022618}
}

@article{Chaki2024universal,
  author = {Chaki, P. and Bhattacharyya, A. and Sen, U.},
  title = {Universal and complete extraction for energy-invariant catalysis in quantum batteries versus no uncorrelated state-invariant catalysis},
  journal = {arXiv preprint},
  year = {2024},
  eprint = {2409.14153},
  archivePrefix = {arXiv},
  primaryClass = {quant-ph}
}

@article{PhysRevLett.85.437,
  title = {Catalysis of Entanglement Manipulation for Mixed States},
  author = {Eisert, J. and Wilkens, M.},
  journal = {Phys. Rev. Lett.},
  volume = {85},
  issue = {2},
  pages = {437--440},
  year = {2000},
   doi = {10.1103/PhysRevLett.85.437}
}

@article{LipkaBartosik2021,
  author = {Lipka-Bartosik, P. and Skrzypczyk, P.},
  title = {All states are universal catalysts in quantum thermodynamics},
  journal = {Phys. Rev. X},
  volume = {11},
  number = {1},
  pages = {011061},
  year = {2021},
  doi = {10.1103/PhysRevX.11.011061}
}

@article{Lie2023,
  author = {Lie, S. H. and Jeong, H.},
  title = {Delocalized and dynamical catalytic randomness and information flow},
  journal = {Phys. Rev. A},
  volume = {107},
  number = {4},
  pages = {042411},
  year = {2023},
  doi = {10.1103/PhysRevA.107.042411}
}

@article{Li2011,
  author = {Li, D. C. and Shi, Z. K.},
  title = {Catalysis of entanglement transformation for 2 $\times$ 2-dimensional mixed states},
  journal = {Quantum Inf. Process.},
  volume = {10},
  number = {5},
  pages = {697--704},
  year = {2011},
  doi = {10.1007/s11128-011-0229-y}
}

@article{Kondra2021a,
  author = {Kondra, T. V. and Datta, C. and Streltsov, A.},
  title = {Catalytic transformations of pure entangled states},
  journal = {Phys. Rev. Lett.},
  volume = {127},
  number = {15},
  pages = {150503},
  year = {2021},
  doi = {10.1103/PhysRevLett.127.150503}
}

@article{Jonathan1999,
  author = {Jonathan, D. and Plenio, M. B.},
  title = {Entanglement-assisted local manipulation of pure quantum states},
  journal = {Phys. Rev. Lett.},
  volume = {83},
  number = {17},
  pages = {3566--3569},
  year = {1999},
  doi = {10.1103/PhysRevLett.83.3566}
}

@article{Ferraro2018,
  author = {Ferraro, D. and Campisi, M. and Andolina, G. M. and Pellegrini, V. and Polini, M.},
  title = {High-power collective charging of a solid-state quantum battery},
  journal = {Phys. Rev. Lett.},
  volume = {120},
  number = {11},
  pages = {117702},
  year = {2018},
  doi = {10.1103/PhysRevLett.120.117702}
}

@article{Gumberidze2022,
  author = {Gumberidze, M. and Kol\'a\v{r}, M. and Filip, R.},
  title = {Pairwise-measurement-induced synthesis of quantum coherence},
  journal = {Phys. Rev. A},
  volume = {105},
  number = {1},
  pages = {012401},
  year = {2022},
  doi = {10.1103/PhysRevA.105.012401}
}

@article{Hu2022,
  author = {Hu, C. K. and Qiu, J. and Souza, P. J. P. and Yuan, J. and Zhou, Y. and Zhang, L. and Chu, J. and Pan, X. and Hu, L. and Li, J. and Xu, Y. and Zhong, Y. and Liu, S. and Yan, F. and Tan, D. and Bachelard, R. and Villas-Boas, C. J. and Santos, A. C. and Yu, D.},
  title = {Optimal charging of a superconducting quantum battery},
  journal = {Quantum Sci. Technol.},
  volume = {7},
  number = {4},
  pages = {045018},
  year = {2022},
  doi = {10.1088/2058-9565/ac80d6}
}

@article{Gumberidze2019,
  author = {Gumberidze, M. and Kol\'a\v{r}, M. and Filip, R.},
  title = {Measurement induced synthesis of coherent quantum batteries},
  journal = {Sci. Rep.},
  volume = {9},
  pages = {19628},
  year = {2019},
  doi = {10.1038/s41598-019-55972-7}
}

@article{Francica2017,
  author = {Francica, G. and Goold, J. and Plastina, F. and Paternostro, M.},
  title = {Daemonic ergotropy: enhanced work extraction from quantum correlations},
  journal = {npj Quantum Inf.},
  volume = {3},
  pages = {12},
  year = {2017},
  doi = {10.1038/s41534-017-0012-8}
}

@article{Downing2024,
  author = {Downing, C. A. and Ukhtary, M. S.},
  title = {Energetics of a pulsed quantum battery},
  journal = {Europhys. Lett.},
  volume = {146},
  number = {1},
  pages = {10001},
  year = {2024},
  doi = {10.1209/0295-5075/acf3b6}
}

@article{Farina2019,
  author = {Farina, D. and Andolina, G. M. and Mari, A. and Polini, M. and Giovannetti, V.},
  title = {Charger-mediated energy transfer for quantum batteries: An open-system approach},
  journal = {Phys. Rev. B},
  volume = {99},
  number = {3},
  pages = {035421},
  year = {2019},
  doi = {10.1103/PhysRevB.99.035421}
}

@article{Bhattacharyya2024,
  author = {Bhattacharyya, A. and Sen, K. and Sen, U.},
  title = {Noncompletely positive quantum maps enable efficient local energy extraction in batteries},
  journal = {Phys. Rev. Lett.},
  volume = {132},
  number = {24},
  pages = {240401},
  year = {2024},
  doi = {10.1103/PhysRevLett.132.240401}
}

@article{Frey2014,
  author = {Frey, M. and Funo, K. and Hotta, M.},
  title = {Strong local passivity in finite quantum systems},
  journal = {Phys. Rev. E},
  volume = {90},
  number = {1},
  pages = {012127},
  year = {2014},
  doi = {10.1103/PhysRevE.90.012127}
}

@article{Sen2021,
  author = {Sen, K. and Sen, U.},
  title = {Local passivity and entanglement in shared quantum batteries},
  journal = {Phys. Rev. A},
  volume = {104},
  number = {3},
  pages = {L030402},
  year = {2021},
  doi = {10.1103/PhysRevA.104.L030402}
}

@article{Brown2016,
  author = {Brown, E. G. and Friis, N. and Huber, M.},
  title = {Passivity and practical work extraction using Gaussian operations},
  journal = {New J. Phys.},
  volume = {18},
  number = {11},
  pages = {113028},
  year = {2016},
  doi = {10.1088/1367-2630/18/11/113028}
}

@article{Skrzypczyk2015,
  author = {Skrzypczyk, P. and Silva, R. and Brunner, N.},
  title = {Passivity, complete passivity, and virtual temperatures},
  journal = {Phys. Rev. E},
  volume = {91},
  number = {5},
  pages = {052133},
  year = {2015},
  doi = {10.1103/PhysRevE.91.052133}
}

@article{Llobet2015,
  author = {Llobet, M. Perarnau and Hovhannisyan, K. V. and Huber, M. and Skrzypczyk, P. and Tura, J. and Ac$\acute{i}$n, A.},
  title = {Most energetic passive states},
  journal = {Phys. Rev. E},
  volume = {92},
  number = {4},
  pages = {042147},
  year = {2015},
  doi = {10.1103/PhysRevE.92.042147}
}

@article{Pusz1978,
  author = {Pusz, W. and Woronowicz, S. L.},
  title = {Passive states and KMS states for general quantum systems},
  journal = {Commun. Math. Phys.},
  volume = {58},
  number = {3},
  pages = {273--290},
  year = {1978},
  doi = {10.1007/BF01614224}
}

@article{Bhattacharjee2021,
  author = {Bhattacharjee, S. and Dutta, A.},
  title = {Quantum thermal machines and batteries},
  journal = {Eur. Phys. J. B},
  volume = {94},
  number = {12},
  pages = {239},
  year = {2021},
  doi = {10.1140/epjb/s10051-021-00299-2}
}

@article{RevModPhys.96.025005,
  title = {Catalysis in quantum information theory},
  author = {Lipka-Bartosik, P. and Wilming, H. and Ng, Nelly H. Y.},
  journal = {Rev. Mod. Phys.},
  volume = {96},
  issue = {2},
  pages = {025005},
  year = {2024},
  month = {Jun},
 doi = {10.1103/RevModPhys.96.025005}
}

@article{Allahverdyan2004,
  author = {Allahverdyan, A. E. and Balian, R. and Nieuwenhuizen, Th. M.},
  title = {Maximal work extraction from finite quantum systems},
  journal = {Europhys. Lett.},
  volume = {67},
  number = {4},
  pages = {565},
  year = {2004},
  doi = {10.1209/epl/i2004-10101-2}
}

@article{Ghosh2020,
  author = {Ghosh, S. and Chanda, T. and Sen(De), A.},
  title = {Enhancement in the performance of a quantum battery by ordered and disordered interactions},
  journal = {Phys. Rev. A},
  volume = {101},
  number = {3},
  pages = {032115},
  year = {2020},
  doi = {10.1103/PhysRevA.101.032115}
}

@article{Barra2019,
  author = {Barra, F.},
  title = {Dissipative charging of a quantum battery},
  journal = {Phys. Rev. Lett.},
  volume = {122},
  number = {21},
  pages = {210601},
  year = {2019},
  doi = {10.1103/PhysRevLett.122.210601}
}

@article{Cavaliere2025,
  author = {Cavaliere, F. and Gemme, G. and Benenti, G. and Ferraro, D. and Sassetti, M.},
  title = {Dynamical blockade of a reservoir for optimal performances of a quantum battery},
  journal = {Commun. Phys.},
  volume = {8},
  number = {1},
  pages = {76},
  year = {2025},
  doi = {10.1038/s42005-025-01993-7}
}

@article{Binder2015,
  author = {Binder, F. C. and Vinjanampathy, S. and Modi, K. and Goold, J.},
  title = {Quantacell: powerful charging of quantum batteries},
  journal = {New J. Phys.},
  volume = {17},
  number = {7},
  pages = {075015},
  year = {2015},
  doi = {10.1088/1367-2630/17/7/075015}
}

@article{Alicki2013,
  author = {Alicki, R. and Fannes, M.},
  title = {Entanglement boost for extractable work from ensembles of quantum batteries},
  journal = {Phys. Rev. E},
  volume = {87},
  number = {4},
  pages = {042123},
  year = {2013},
  doi = {10.1103/PhysRevE.87.042123}
}

@incollection{Campaioli2018,
  author = {Campaioli, F. and Pollock, F. A. and Vinjanampathy, S.},
  title = {Quantum Batteries},
  booktitle = {Thermodynamics in the Quantum Regime},
  editor = {Binder, F. and Correa, L. A. and Gogolin, C. and Anders, J. and Adesso, G.},
  publisher = {Springer},
  pages = {207--225},
  year = {2018},
  doi = {10.1007/978-3-319-99046-0_11}
}

@article{Andolina2019,
  author = {Andolina, G. M. and Farina, D. and Mari, A. and Pellegrini, V. and Giovannetti, V. and Polini, M.},
  title = {Charger-mediated energy transfer in exactly solvable models for quantum batteries},
  journal = {Phys. Rev. B},
  volume = {99},
  number = {3},
  pages = {035421},
  year = {2019},
  doi = {10.1103/PhysRevB.99.035421}
}

@article{Le2018,
  author = {Le, T. P. and Levinsen, J. and Modi, K. and Parish, M. M. and Pollock, F. A.},
  title = {Spin-chain model of a many-body quantum battery},
  journal = {Phys. Rev. A},
  volume = {97},
  number = {2},
  pages = {022106},
  year = {2018},
  doi = {10.1103/PhysRevA.97.022106}
}

@article{Santos2019,
  author = {Santos, A. C. and \c{C}akmak, B. and Campbell, S. and Zinner, N. T.},
  title = {Stable adiabatic quantum batteries},
  journal = {Phys. Rev. E},
  volume = {100},
  number = {3},
  pages = {032107},
  year = {2019},
  doi = {10.1103/PhysRevE.100.032107}
}

@article{Arrachea2023,
  author = {Arrachea, L.},
  title = {Energy dynamics, heat production and heat--work conversion with qubits: toward the development of quantum machines},
  journal = {Rep. Prog. Phys.},
  volume = {86},
  number = {3},
  pages = {036501},
  year = {2023},
  doi = {10.1088/1361-6633/acb06b}
}

@article{PhysRevA.90.022110,
  title = {Quantum regression theorem and non-Markovianity of quantum dynamics},
  author = {Guarnieri, G. and Smirne, A. and Vacchini, B.},
  journal = {Phys. Rev. A},
  volume = {90},
  issue = {2},
  pages = {022110},
  year = {2014},
  doi = {10.1103/PhysRevA.90.022110}
}

@article{Quach2020,
  author = {Quach, J. Q. and McGhee, K. E. and Ganzer, L. and Dominici, L. and Savona, V. and Li, F. and Liew, T. C. H. and Berloff, N. G. and Galopin, E. and Lemaitre, A. and Bloch, J. and Amo, A.},
  title = {Superabsorption in an organic microcavity: Toward a quantum battery},
  journal = {Sci. Adv.},
  volume = {6},
  number = {14},
  pages = {eaaz4487},
  year = {2020},
  doi = {10.1126/sciadv.aaz4487}
}

@article{Seah2021,
  author = {Seah, S. and Nimmrichter, S. and Grimmer, D. and Santos, J. P. and Scarani, V. and Landi, G. T.},
  title = {Quantum thermodynamics with local master equations},
  journal = {Phys. Rev. Lett.},
  volume = {127},
  number = {10},
  pages = {100601},
  year = {2021},
  doi = {10.1103/PhysRevLett.127.100601}
}

@article{PhysRevA.107.042419,
  title = {Catalysis in charging quantum batteries},
  author = {Rodriguez, R. R. and Ahmadi, B. and Mazurek, P. and Barzanjeh, S. and Alicki, R. and Horodecki, P.},
  journal = {Phys. Rev. A},
  volume = {107},
  issue = {4},
  pages = {042419},
  year = {2023},
  doi = {10.1103/PhysRevA.107.042419}
}

@article{Rossini2019,
  author = {Rossini, D. and Andolina, G. M. and Polini, M.},
  title = {Many-body localized quantum batteries},
  journal = {Phys. Rev. B},
  volume = {100},
  number = {11},
  pages = {115142},
  year = {2019},
  doi = {10.1103/PhysRevB.100.115142}
}

@article{Blatt2012,
  author = {Blatt, R. and Roos, C. F.},
  title = {Quantum simulations with trapped ions},
  journal = {Nat. Phys.},
  volume = {8},
  pages = {277--284},
  year = {2012},
  doi = {10.1038/nphys2252}
}

@article{Blais2021,
  author = {Blais, A. and Grimsmo, A. L. and Girvin, S. M. and Wallraff, A.},
  title = {Circuit Quantum Electrodynamics},
  journal = {Rev. Mod. Phys.},
  volume = {93},
  number = {2},
  pages = {025005},
  year = {2021},
  doi = {10.1103/RevModPhys.93.025005}
}

@article{Gu2017,
  author = {Gu, X. and Kockum, A. F. and Miranowicz, A. and Liu, Y. x. and Nori, F.},
  title = {Microwave photonics with superconducting quantum circuits},
  journal = {Phys. Rep.},
  volume = {718},
  pages = {1--102},
  year = {2017},
  doi = {10.1016/j.physrep.2017.10.002}
}

@article{Zhao2025nspin,
  title = {Non-Markovian {$N$}-spin chain quantum battery in thermal charging process},
  author = {Zhao, S. C. and Zhao, Z. R. and Zhuang, N. Y.},
  journal = {Phys. Rev. E},
  volume = {112},
  issue = {2},
  pages = {024129},
  year = {2025},
  publisher = {American Physical Society},
  doi = {10.1103/xqtv-qbyk},
  url = {https://link.aps.org/doi/10.1103/xqtv-qbyk}
}
\bibliographystyle{unsrt}%%{apsrev4-2,unsrturl, abbrv, acm, alpha, apalike, ieeetr, plain, siam , unsrt}

\end{document}